# Quadruplets of exceptional points and bound states in the continuum in dielectric rings

Nikolay Solodovchenko[1], Kirill Samusev[1,2] and Mikhail Limonov[1,2]

[1] Department of Physics and Engineering, ITMO University, St. Petersburg, 197101, Russia

[2] Ioffe Institute, St. Petersburg 194021, Russia

**In photonics, most systems are non-Hermitian due to radiation into open space and material losses. At the same time, non-Hermitianity defines a new physics, in particular, it gives rise to a new class of degenerations called exceptional points, where two or more resonances coalesce in both eigenvalues and eigenfunctions. The point of coalescence is a square root singularity of the energy spectrum as a function of interaction parameter. We investigated analytically and numerically the photonic properties of a narrow dielectric resonator with a rectangular cross section. It is shown that the exceptional points in such a resonator exist in pairs, and each of the points is adjacent in the parametric space to a bound state in the continuum, as a result of which quadruples of singular photonic states are formed. We also showed that the field distribution in the cross section of the ring is a characteristic fingerprint of both the bound state in the continuum and the exceptional point.**

Our world is arranged in such a way that most systems are not closed, interact with each other and are described by non-Hermitian physics. The non-Hermiticity provides rich topological properties that often have no analogue in Hermitian structures. As a rule, a non-Hermitian eigenvalue problem does not have an orthogonal set of eigenvectors; moreover, eigenvectors can be collinear. A point in the parameter space at which non-Hermitian degeneracy is observed, i.e. at least two eigenvalues and eigenvectors coalesce, is called the exceptional point (EP) [1, **2**]. At EPs, two energy levels are connected by a square-root branch point; moreover they are the values of one analytic function on two different Riemann eigenvalue sheets [3-12]. In this paper, we study the two-level problem, which is the simplest case of a non-Hermitian system. In general, EPs appear in various systems with spatially discrete or continuous degrees of freedom of multiple dimensionalities and also can be considered as a critical point, near which there is a transition from strong coupling to weak coupling [5]. If the system is described by more than two eigenvalue surfaces, then it is possible that more than two surfaces simultaneously collapse at the same point, creating a higher-order EP [13,14]. In particular, a third-order EP occures when three eigenvalues simultaneously coalesce and the square-root dependence of the eigenvalues around the exceptional point is replaced by a cube root.

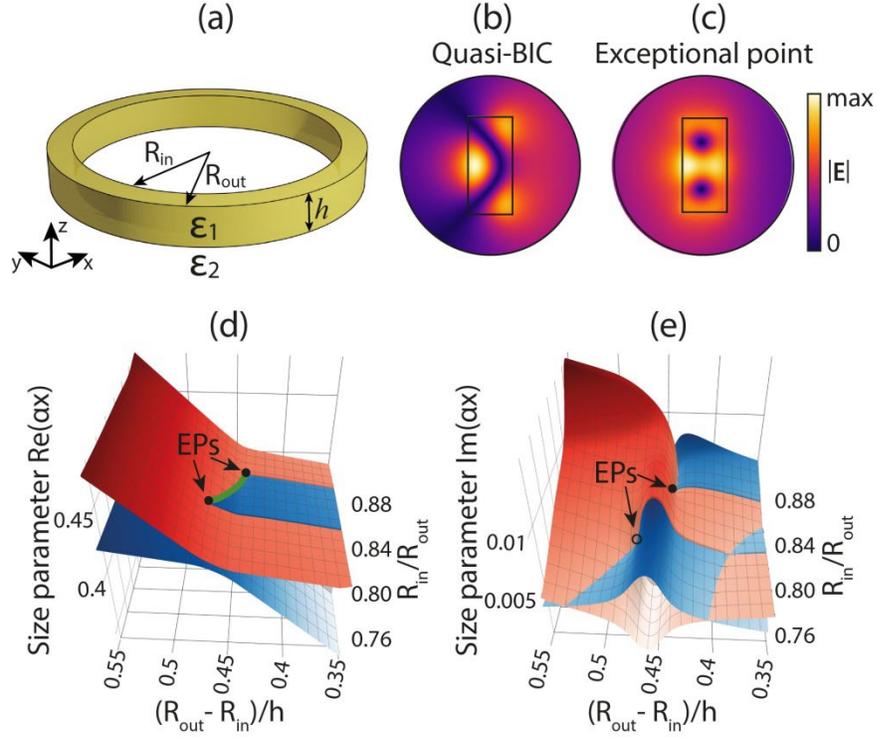

**Fig. 1 | Exceptional points and bound states in the continuum in dielectric ring. a,** Dielectric ring with permittivity $\varepsilon_1$, inner radius $R_{in}$, outer radius $R_{out}$, and height h placed in vacuum ($\varepsilon_2 = 1$). **b,c,** Calculated pattern of the electric field amplitude |E| inside the resonator and around it for the qBIC and the EP. **d, e,** Perspective view of the complex square-root topology of calculated eigenvalue surfaces $Re(\alpha x)$ and $Im(\alpha x)$ with a branch point singularity at two EP in the parameter space ($R_{out}-R_{in}$)/h, $R_{in}/R_{out}$. In **d**, two EPs are connected by a Fermi arc (green line). In **e**, one of the EPs is hidden in this perspective by the bend of the Riemann sheet and depicted by a ring.

For a narrow dielectric ring resonator (RR) with a rectangular cross section, the result of plotting the full dependence of both eigenvalues of the non-Hermitian Hamiltonian on structural parameters ($R_{out}-R_{in}$)/h, $R_{in}/R_{out}$ demonstrates paired EPs, Fig 1 (d,e). Ring parameters are: ($R_{out}-R_{in}$)/h=0.472, $R_{in}/R_{out}$=0.8005, $\varepsilon_1$=80, the size parameter $\alpha x = 0.418$. Paired EP generate a distinct double-Riemann-sheet topology in the complex band structure, which leads to bulk Fermi arcs [7]. Figure 1(c) shows an unusual shape of the field distribution pattern in the cross section of the dielectric narrow RR. Such a field distribution, which is not observed for other photonic resonances of the dielectric ring, can be considered as a fingerprint of the EP.

EPs are observed in both quantum mechanical and classical problems. EPs are involved in quantum chaos and quantum phase transition, they produce impressive effects in a specific time dependence and multichannel scattering [5,8,11,15]. In various systems, isolated EPs in the parameter space [3, 16-18] and continuous rings of EPs in the momentum space [19] were previously studied. In optics, active or passive systems with EPs exhibit various exotic properties, such as nanoscale sensing [20], laser mode selection [21], electromagnetically induced transparency at a chiral EP [22], asymmetric mode switching [17], directional omni-polarizer [23], directional total absorption [24], and enhancement of Sagnac sensitivity [25] have been proposed and/or demonstrated.

EPs are closely related to the phenomenon of level repulsion [2], which has been originally explored in the context of quantum chaos [26]. In photonics, level repulsion is of interest because it indicates strong coupling and hybridization between states and typically occurs near an exceptional point in the real or complex parameter space [8]. One of the manifestations of level repulsion is associated with bound states in the continuum that arise in accordance with the Friedrich–Wintgen model [27]. The history of BICs begins with the work of von Neumann and Wigner [28], who mathematically modeled a quantum system that has bound states above the continuum threshold. After almost 60 years, it was demonstrated that a similar phenomenon occurs in Maxwell's theory [29], after which the number of publications on this topic has continuously increased. BIC with an infinite quality factor is a mathematical model, and in real structures of finite sizes, quasi-BIC (qBIC) with a large but finite $Q$-factor is studied both theoretically and experimentally. The Friedrich–Wintgen's qBIC occurs when two non-orthogonal modes are coupled to the same radiation channel, interfere in the near field, and the avoided crossing arises in the parametric space [30-33]. Recently, the Friedrich-Wintgen model has been successfully utilized to interpret qBIC in dielectric cylinders [34] and rings [35]. For a single dielectric cylinder, qBIC occurs when two eigenmodes with different polarizations, associated with Mie-like resonances between the side walls and Fabry- Pérot – like resonances between the top and bottom walls of the cylinder, form an avoided crossing region in the strong coupling regime. These modes are approximately orthogonal inside the cylinder and interfere mainly in the near field zone of the cylinder [14], realize qBIC, while the all-dielectric resonator demonstrates extremely high values of the $Q$ factor. In many cases the appearance of qBIC was accompanied by the presence of the Fano resonance [29, 34, 35]. In particular, the link between the physics of Fano resonances [36-38] and qBICs excited in individual high-index dielectric nanoparticles has been experimentally demonstrated in Ref. 39.

In this work, we have investigated the behavior of another pair of resonant modes $TE_{0,2,0}$ and $TE_{0,1,2}$ ($m$=0), which also form a q-BISs according to the Friedrich–Wintgen model [28], but the study is carried out up to very narrow rings $R_{in}/R_{out}$ = 0.95. As a result, we discovered two quadruplets (EP + qBIC) of singular photonic states: it turned out that each exceptional point of the pair closely neighbors in the coordinate space with a bound state in the continuum, thus forming a quadruple of singular states of the $TE_{0,2,0}$ and $TE_{0,1,2}$ interfering photonic modes. We demonstrate two photonic quadruples, one in the $R_{in}/R_{out}$ = 0.80-0.84 region and the other in the $R_{in}/R_{out}$ = 0.92-0.93 region.

## Calculation Methods

Three methods were used to calculate the eigenvalues of a dielectric RR with a rectangular cross section: COMSOL Multiphysics, Resonant state expansion (RSE), and Temporal coupling modes theory (TCMT).

The COMSOL Multiphysics program allows using the optical module to find eigenvalues (resonance frequencies) and eigenfunctions (electromagnetic field distributions), as well as the scattering cross section (SCS) $\sigma_{sca}$. In 3D RRs the eigenfunctions can be characterized by the azimuthal ($m$), radial ($r$), and axial ($z$) mode indices, forming ordered triple ($m, r, z$). Since Maxwell's equations are scaled in the absence of dispersion, the defining geometric size (for example, the outer radius $R_{out}$) can be chosen arbitrarily. The inner radius of the RR varied over a wide range from $R_{in}/R_{out}$ = 0 (cylinder) to $R_{in}/R_{out}$ = 0.95 (narrow ring). The dielectric constant of the RR was chosen to be 80, when the resonance effects are most pronounced, which corresponds, for example, to high-index ceramics in the microwave range. To calculate the eigenvalues and eigenfunctions in COMSOL, the "Eigenfrequency" mode was used, while the resonator was surrounded by PML (perfect matched layer), and the incident wave was absent. The scattering cross section was calculated only for one partial harmonic of the corresponding symmetry of the considered modes, m=0. The scattering cross section $\sigma_{sca}$,0 was normalized to S = $2R_{out}h$ (see Supplementary Section A for more details).

Another approach to find the eigenvalues of the dielectric RR used in this work is RSE, which is an exact method of perturbation theory [40,41]. As a set of basis functions, we took the eigenvectors of a dielectric sphere with the same permittivity $\varepsilon_a^{(0)}$ with frequencies $kR_{out,sphere}$ < 20 and orbital angular momentum l <60, which is sufficient for convergence of 99.5% of the real part of the frequency even for a narrow ring, geometrically distant from the sphere. The problem of finding the eigenvalues of the RR within the framework of this method is reduced to a matrix equation [see. Supplementary Materials], where the perturbation function changes the permittivity of the sphere, turning it into a ring, and the perturbation coefficients are found using the eigenfunctions of the sphere. Note that the problem is non-Hermitian due to the outgoing wave. Consequently, the eigenvectors grow exponentially at large distances, and their correct normalization deviates from the standard [40, 41] (See Supplementary Section A for more details).

The mechanism of the appearance of the Friedrich – Wintgen's qBIC suggests that this effect is close to the appearance of EP. Indeed, in this case, qBICs arise as a result of the interaction of two eigenmodes of the structure, which can lead to level anticrossing [see Supplementary Materials], but, under certain structure parameters, can also lead to level crossing $Re_1(\alpha x) = Re_2(\alpha x)$, which, in general, corresponds to a transition from a strong-coupling to a weak-coupling regimes [35]. However, an even more intriguing case is possible, when both real and imaginary parts of the eigenvalues coalesce, and an EP is achieved: $Re_1(\alpha x) = Re_2(\alpha x)$ and $Im_1(\alpha x) = Im_2(\alpha x)$. Therefore, all essential aspects of the Friedrich – Wintgen's qBICs and EPs can be illustrated on an elementary level with the same two-level model. According to the temporal coupled-mode theory [1], when two resonances are eigenmodes of one resonator and are coupled to the same radiation channel, the amplitudes evolve with the Hamiltonian [2-5]:

$$H = \begin{pmatrix} \omega_1 & k \\ k & \omega_2 \end{pmatrix} - i \begin{pmatrix} \gamma_1 & \sqrt{\gamma_1 \gamma_2} \\ \sqrt{\gamma_1 \gamma_2} & \gamma_2 \end{pmatrix}. \quad (1)$$

We consider the case when the two uncoupled resonances can have different resonance frequencies, $\omega_{1,2}$, and different radiative damping rates, $\gamma_{1,2}$. Here, $\kappa$ is the internal (near-field) coupling between the two resonators that radiate into the same channel, and therefore the interference of radiation gives rise to the via-the-continuum coupling term $\sqrt{\gamma_1 \gamma_2}$. The condition for the occurrence of qBIC according to Friedrich – Wintgen's model [27] has the form:

$$k_{\text{qBIC}} = \frac{(\omega_1 - \omega_2)\sqrt{\gamma_1 \gamma_2}}{(\gamma_1 - \gamma_2)} \quad (2)$$

Satisfying this condition, one can achieve a significant suppression of the total radiation losses of one of the modes. The resulting mode will be qBIC since it is formed according to the Friedrich − Wintgen`s mechanism, but some of the radiation was not suppressed due to coupling with several radiation channels, which can be taken into account as an additional imaginary part to the coupling coefficient. As a result, the radical expression is extracted and we obtain expressions for two modes one of which is mostly real and becomes a qBIC.

The situation with the EP corresponds to the equality of the real and imaginary parts of the frequency. In contrast to the qBIC, the EP requires the setting of two parameters at once instead of one. Specifically, in the case of a ring, the parameters may be the height and radius of the inner hole. To observe the EP, two conditions must be met:

$$\begin{cases} (\omega_1 - \omega_2) = -\frac{4\sqrt{\gamma_1 \gamma_2} k_{EP}}{(\gamma_1 - \gamma_2)} & (3) \\ |k_{EP}| = \pm(\gamma_1 - \gamma_2)/2 & (4) \end{cases}$$

When $|\kappa| < |\gamma_1 - \gamma_2|/2$ corresponds to weak coupling and the modes intersec, and $|\kappa| > |\gamma_1 - \gamma_2|/2$ represents strong coupling and the modes anti-cross. Condition for $\kappa_{EP}$ resembles condition $\kappa_{q-BIC}$ for the qBIC, but with inverted terms of the radiative damping rates, $\gamma_{1,2}$ and have different signs, so are observed on different sides with respect to the intersection of the real parts of the uncoupling modes.

In the case of interaction of two modes in a single dielectric resonator, described by Hamiltonian (3), EPs always appear in pairs, which is demonstrated by formula (4). Paired EPs [7, 9, 42] are connected in parameter space by an open arc, known as a bulk Fermi arc, along which the resonant frequencies of the two modes are degenerate [green line in Fig. 1(d)], but have different radiative damping rates, Fig. 1(e). Between two EP, the coupling coefficient is $|\kappa| < |\gamma_1 - \gamma_2|/2$. Using Hamiltonian (3) eigenvalue functions of COMSOL Multiphysics can be fitted and coupling coefficients can be defined (see Supplementary Section C for more details).

## Results and discussion: two quadruples of singular points

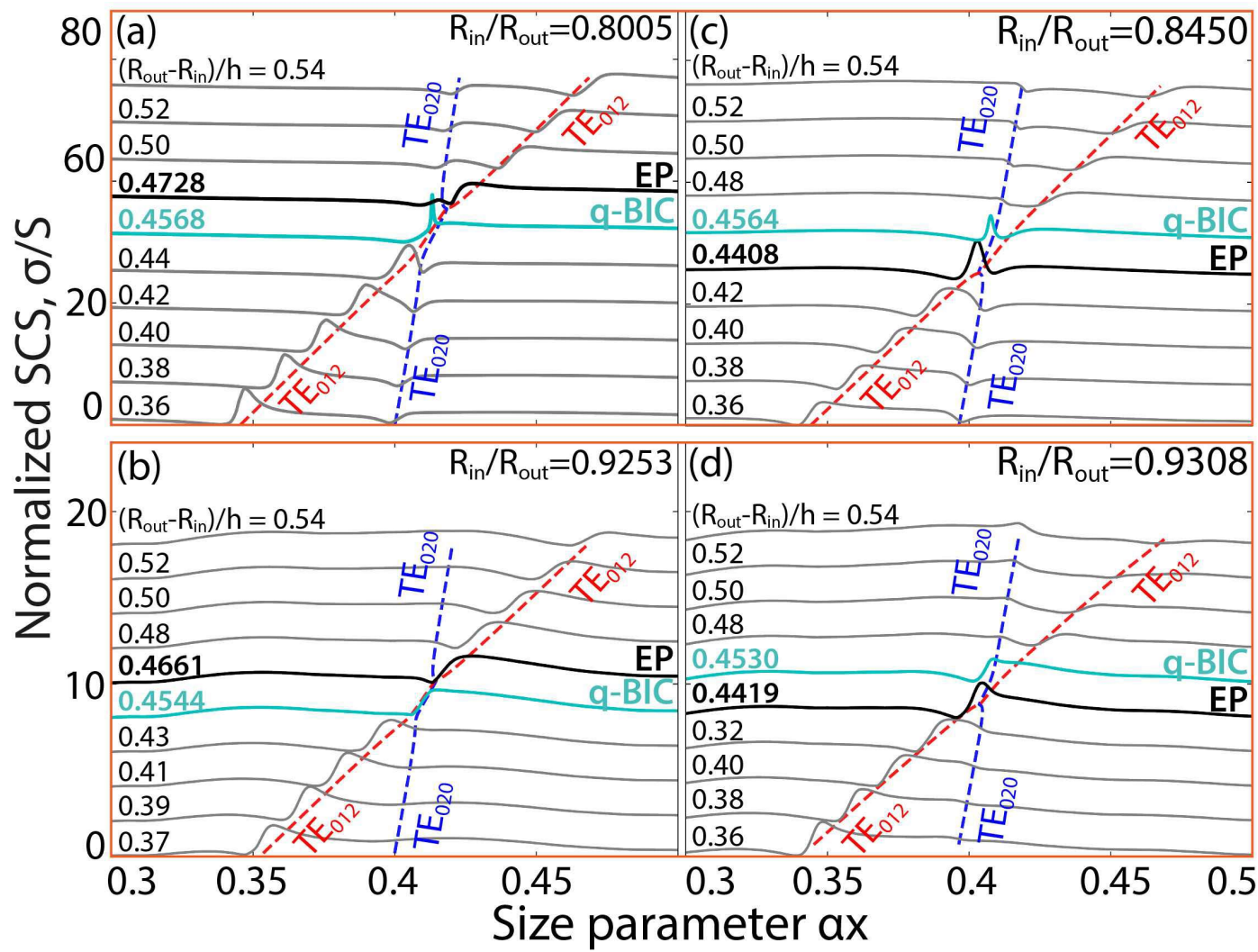

**Fig. 2 | Calculated spectra of the normalized SCS for dielectric RRs for two (EP + qBIC) quadruplets in yellow frames.** The spectra are presented on the aspect ratio $(R_{out} - R_{in})/h$. The first quadruplet is shown in the top row, the second quadruplet in the bottom row, the Fabry-Pérot-type $TE_{0,1,2}$ and Mie-type $TE_{0,2,0}$ modes interfere. SCS spectra are given in the range of $0.36 \leq R_{out}/h \leq 0.54$ with a step of 0.02. Curves are shifted vertically by 6 a.u for $R_{in}/R_{out}=0.8005, 0.8450$ and by 2 a.u for $R_{in}/R_{out}=0.9253, 0.9308$. The spectra highlighted with a thick black line correspond to EP, and the spectra highlighted with a thick purple line correspond to qBIC. Dotted lines are only a guide for the eyes. TE-polarized incident wave. $\varepsilon_1 = 80$, $\varepsilon_2 = 1$.

Figure 3, which shows the calculated normalized scattering cross-section (SCS/S, S=$2R_{out}$ h) spectra of dielectric RRs in the spectral region of the $TE_{0,1,2}$ and $TE_{0,2,0}$ resonances. The results obtained make it possible to trace the emergence and transformation of both qBICs and EPs. For four fixed value of $R_{in}/R_{out}$, the SCS of the dielectric ring resonator ware calculated as a function of its aspect ratio $(R_{out} - R_{in})/h$ upon excitation by a plane TE-polarized wave. With this value, firstly, the resonance scattering spectra have narrow peaks that are convenient for treatment and interpretation. For generality, when demonstrating the results we use the normalized size parameter $\alpha x = kR_{out}\left(1 - \frac{R_{in}}{R_{out}}\right)$ being a product of the wavenumber $k$ and outer radius $R_{out}$.

We have previously demonstrated that in dielectric cylinders and rings there are two types of modes with different behavior depending on the aspect ratio $R_{out}/h$ [34,35]. Modes of the first type are formed mainly due to reflection from the side wall, are associated with the Mie resonances of an infinite cylinder, and, accordingly, have a slight frequency shift with a change in the length of the resonator. Modes of the second type are formed mainly due to reflection from two parallel faces of a cylinder or ring; they are similar to Fabry-Pérot modes and demonstrate a significant shift in the spectra with a change in the resonator length. Due to different spectral shifts, the $TE_{0,2,0}$ Mie-type and $TE_{0,1,2}$ Fabry-Perot-type modes can have the same frequencies at some points in the parameter space, Figure 3. Since the modes have the same azimuthal indices $m=0$, they interact, and one of them turns into qBIC ($\gamma \to 0$), which is described by the Friedrich-Wintgen model [27]. The qBIC line in the scattering spectra can sharply narrow and, because of this, disappear both in the calculated and experimental spectra. In particular, the qBIC line in the calculated scattering spectra (Comsol Multiphysics) on a dielectric cylinder with a permittivity $\varepsilon = 80$ was traced only up to a Q factor of $10^4$ [44]. In the case of weak coupling, the lines corresponding to the $TE_{0,2,0}$ and $TE_{0,1,2}$ modes intersect in the scattering spectra, Figure 3. At the same time, it turns out that in narrow dielectric RRs near the point where the qBIC appears, not only the real parts of the eigenvalues, but also the imaginary parts can coalesce [$Re_1(\alpha x) = Re_2(\alpha x)$, $Im_1(\alpha x) = Im_2(\alpha x)$], which leads to EPs that are formed in pairs, as follows from the analytical analysis (Eq.12), and from numerical calculations, Figs.1,4.

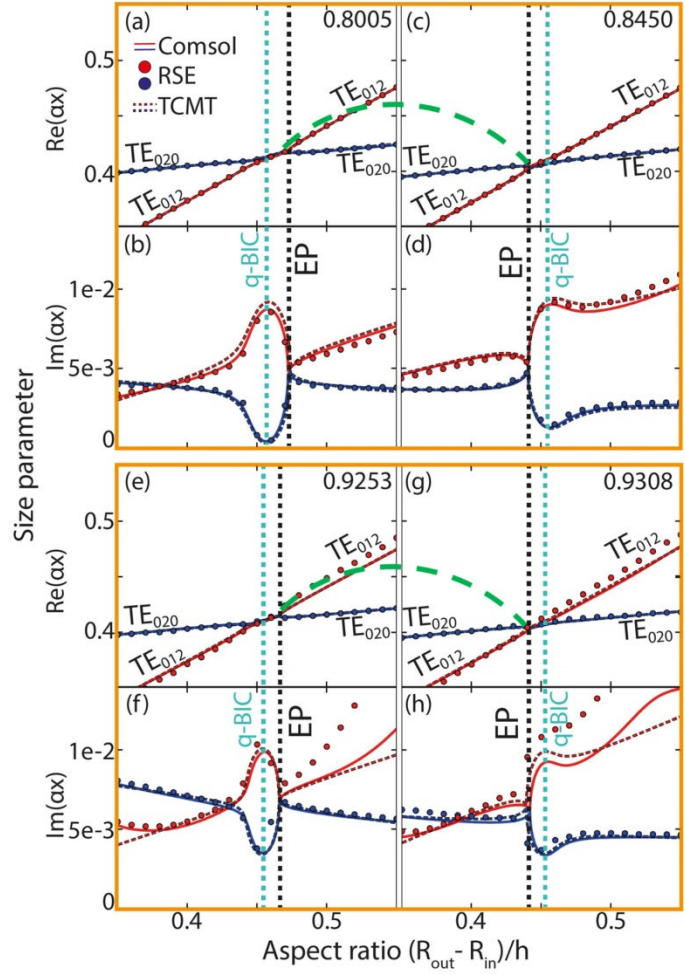

**Fig. 3 | Eigenvalues Re(αx) and Im(αx) for two (EP + qBIC) quadruplets, in yellow frames.** Calculations for dielectric RRs with a rectangular cross section depending on the aspect ratio $(R_{out} - R_{in})/h$ and the parameter $R_{in}/R_{out}$, indicated at the top of the panels (a,c,e,g). Panels (a-d) refer to the resonances of the first quadruplet, panels (e-h) to the resonances of the second quadruplet. (a, c, e, g) Calculated real part of frequencies of the Fabry-Pérot-type $TE_{0,1,2}$ and the Mie-type $TE_{0,2,0}$ modes in qBIC and EP regimes as a function of the aspect ratio. (b, d, f, h) Imaginary part of frequencies evolution demonstrating the presence of both qBIC and EP. The thin vertical dotted line indicates the position of the qBIC, and the thick line indicates the position of the EP. The green dashed lines indicate the "symbolic" Fermi arcs that connect EPs in the parametric space. The results of calculations by three methods: continuous curves – COMSOL, dotted lines – TCMT, circles - RSE. The dielectric permittivity of RRs is ε=80. The RRs are placed in the vacuum, ε =1.

Figure 3 presents the main result of this work — the observation of two pairs of EPs, with each EP located in close proximity to another singular point, namely, to the qBIC. Indeed, at this singular point, the modes intersect, which means a weak coupling regime, however, in contrast to EP, one of the modes has a local maximum of the Q factor, and the other a local minimum of the Q factor, which is described by Eqs. 8(a,b). Thus, the pair EPs and the pair qBICs form a quadruplet (EP + qBIC) of connected singular points in the parametric space. In a very narrow dielectric ring, we found two quadruplets, and the singular points of the lower quadruplet in Fig. 4(e-h) are especially close to each other, differing in the normalized width of the ring $R_{in}/R_{out}$ by only 0.0055.

It should be noted that the results obtained by three different methods are in good agreement. In particular, in many cases the results were so close that the lines in Fig. 4 completely coincided and the dotted line that represents the TCMT results becomes indistinguishable against the background of the solid lines representing the results of the calculations in COMSOL.

Figure 4 shows a significant analogy between the two quadruplets. In particular, a pair of singular points in quadruplets with a smaller $R_{in}/R_{out}$ parameter (0.8005 and 0.9253) is located in the qBIC - EP sequence in terms of aspect ratio $(R_{out} - R_{in})/h$, and a pair of singular points in quadruplets with a large $R_{in}/R_{out}$ (0.8450 and 0.9308) is located in the opposite sequence EP - qBIC.

However, there is also a difference between quadruplets, which is expressed in the form of a contour for the $TE_{0,2,0}$ Mie-type resonance in the scattering spectra of two quadruplets. Figure 3 shows that in the spectra of the first quadruplet this resonance has an inverted line shape in the form of a dip (circled in green oval), while the line of the second quadruplet is not inverted. The asymmetric line shape in such scattering spectra is determined by the Fano interference between narrow resonant lines and a broad background [36-38] (see Supplementary Section D for more details).

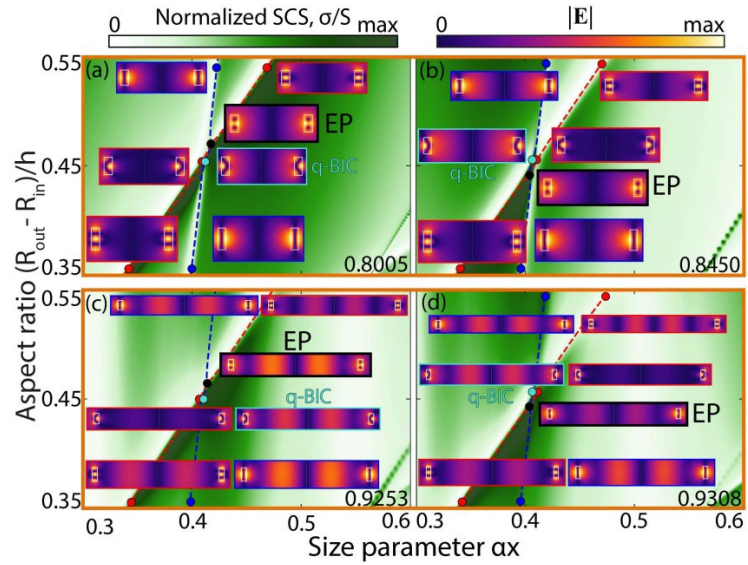

**Fig. 4 | Calculated spectra of the normalized SCS and distribution of the electric field amplitude for two quadruples (EP + qBIC) in yellow frames.** The calculations were carried out depending on the aspect ratio $(R_{out}-R_{out})/h$ and size parameter $\alpha x$. The calculations are carried out with the step of $(R_{out}-R_{out})/h$=4e-4 (501 spectra in total). TE-polarized incident waves. Inserts: The calculated field patterns (total electric field amplitude $|E|$) in side view of RR for the high-frequency and low-frequency branches of the interacting pair of modes $TE_{0,1,2}$ and $TE_{0,2,0}$. The field distributions are shown for four $(R_{out}-R_{out})/h$ values: for EP in black frame, for qBIC in lilac frame and for two values away from the intersection region, with the high frequency field pattern in the red frame and the low frequency field pattern in the blue frame. The colors of the circles correspond to the colors of the frames. The sections of the ring are circled with white rectangles. The dashed lines show the real part of frequencies. TE-polarized incident wave. $\varepsilon_1 = 80$, $\varepsilon_2 = 1$. Supplementary section E presents eight field distribution patterns (four each for EP and qBIC) on an enlarged scale.

Fig. 4 presents color maps of the normalized SCS for two (EP + qBIC) quadruplets of singular points in the parametric space of aspect ratio $(R_{out}-R_{out})/h$ and size parameter $\alpha x$. This map corresponds to the spectra presented in Fig. 2, however, calculated with a much smaller step in the parameter $(R_{out}-R_{out})/h$. The difference in the behavior of Mie-type and Fabry-Perot-type modes is qualitatively clear. Away from the branch intersection region, the Fabry-Perot resonant modes are determined by the top and bottom faces of the ring and the field varies in height, while the Mie resonances are determined by the side faces of the ring and the field changes in the plane of the ring. However, in the region where the branches intersect, the field patterns change completely. The electric field amplitude $|E|$ of the qBIC is determined by the combination of the Fabry-Perot and Mie resonance fields, which we observed earlier in the case of a cylinder [44] and a ring with a small inner radius [35]. As for EP, we observe for the first time a surprising field distribution, which consists of two maxima of $|E|$ located symmetrically in the cross' section of the ring along the vertical axis. On an enlarged scale, this field distribution is shown in Fig. 1(b), it is not an obvious combination of fields corresponding to Mie and Fabry-Perot resonances, and can be considered as a characteristic fingerprint of EP.

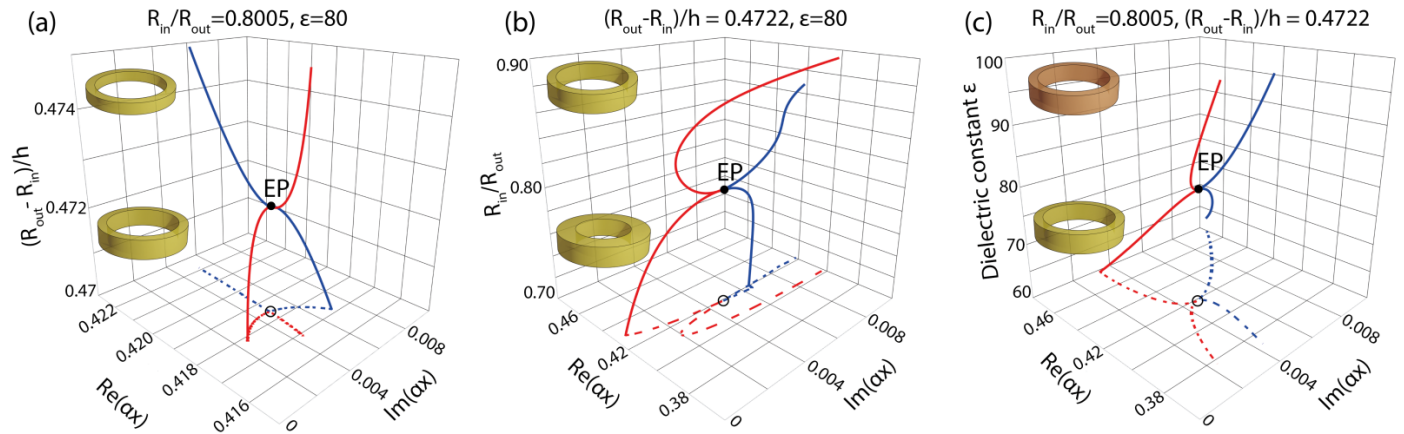

**Fig. 5 | Dispersion of the photonic branches of a dielectric RR with a rectangular cross section in the EP region.** Dependences on the height (a), thickness (b) and permittivity (c) in the coordinate space Re(αx), Im(αx).

The final Fig. 5 shows the dispersion of the photonic branches of a rectangular dielectric RR in the EP region with a change in three main parameters: normalized height (a), thickness (b) and permittivity (c) in the coordinate space Re( αx), Im(αx). The projections of branches are also shown on the plane Re(αx), Im(αx), which give an idea of the dispersion of branches in three-dimensional space. In particular, due to the projection in Fig. 5(a) it becomes clear that when the aspect ratio $(R_{out}-R_{out})/h$ changes at EP, both branches rotate 90 degrees. In this case, the rotation is not observed when the parameter $R_{in}/R_{out}$ and the permittivity ε change, so the change in the height of the ring has the most dramatic effect on the character of the photonic system behavior of the dielectric ring in the EP.

## Conclusions

In summary, in the spectra of dielectric rings, we theoretically discovered quadruplets of singular points formed by two EPs connected by a Fermi arc and two qBICs, each of which is adjacent to one of the EPs. The 2(EP + qBIC) quadruplets were observed by three different methods, the results of which agree perfectly. The appearance of quadruplets is determined by the interaction of two photonic branches, which can anticrossing or intersect in the parametric space during the transition from the strong to the weak coupling regime, which is described by the Friedrich – Wintgen's model. A dielectric ring is an ideal object for modeling quadruplets due to the ability to arbitrarily change the shape of a rectangular cross section, that is, to precisely scan the areas of intersection of axial and radial Fabry-Pérot – like resonances along the height or width of the ring. The key role is played by the internal hole as an additional degree of freedom, which allows one to change the mode coupling coefficient and observe EPs. An important part of the work was the study of the electromagnetic field distribution in the ring and inside it. For EP and qBIC, the distributions have a characteristic form, which can be considered as photonic fingerprints of the corresponding resonances. We note the field in EPs, which corresponds to two zero-intensity "cords" inside the ring. Moreover, we have demonstrated the regimes of electric field concentration inside a narrow dielectric ring at the incidence of a plane electromagnetic wave. Observation of pairs (EP + qBIC) will reveal the topological physics of bound states in the continuum in dielectric resonators in general, as well as create new optoelectronic devices based on resonant topological effects in dielectric rings, expanding the possibilities of sensing, filtering, switching, harmonic generation.

**Acknowledgements**
NS acknowledge support from the Foundation for the Advancement of Theoretical Physics and Mathematics "BASIS" (Russia), KS acknowledges the financing of the project by the Ministry of Science and Higher Education of the Russian Federation (Project 075-15-2021-589), ML acknowledges the financial support from the Russian Science Foundation (project 23-12-00114).
.